\documentclass[a4paper,12pt]{article}
\usepackage{amsmath}
\usepackage{accents}
\usepackage[english]{babel}
\usepackage[usenames]{color}
\usepackage{colortbl}

\begin{document}

\centerline {\bf Geometry of a two-spin quantum state in evolution}
\medskip
\centerline {A. R. Kuzmak$^1$, V. M. Tkachuk$^2$}
\centerline {\small \it E-Mail: $^1$andrijkuzmak@gmail.com, $^2$voltkachuk@gmail.com}
\medskip
\centerline {\small \it Department for Theoretical Physics, Ivan Franko National University of Lviv,}
\medskip
\centerline {\small \it 12 Drahomanov St., Lviv, UA-79005, Ukraine}

{\small

We study the quantum evolution of a two-spin system described by the isotropic Heisenberg Hamiltonian in the external magnetic field.
It is shown that this evolution happens on a two-parametric closed manifold. The Fubini-Study metric of this manifold is obtained.
It is found that this is the metric of the torus. The entanglement of the states which belong to this manifold is investigated.

\medskip

}

\section{Introduction\label{sec1}}

It is well known that quantum theory can be formulated in the geometrical
language \cite{GFofQM, GAQMQE, AGAtoQM, EMCCR}. The geometrical methods are successfully applied to the consideration of the unitary
evolution of quantum states. This is because the states of a quantum system are represented by rays in a complex Hilbert space that in turn
leads to a geometrical formulation of the postulates of quantum mechanics.

For understanding of the dynamics of a quantum system it is useful
to investigate manifolds which contain all states that can be
reached during this dynamics. For instance, the whole state space
of a two-level system (qubit) can be represented by a 2D sphere
called the Bloch sphere. Then the trajectory of quantum evolution
between two states is a curve between two points on this sphere
(see, for instance, \cite{FSM, FSM2, ref1, ref2, ref3}). Using geometrical approach in quantum mechanics one can often find solutions
to many problems in a simple way. For example,
the problem of finding the Hamiltonian which provides time-optimal evolution between two states was solved
using symmetry properties of the quantum state space \cite{OHfST}.
Another interesting problem, which was solved in a similar way, is the Zermelo navigation problem \cite{ZNP1, ZNP2, ZNP3, ZNP4, ZNP5}.
Also, the quantum brachistochrone problem for an arbitrary spin in a magnetic field
was solved using geometrical properties of rotational manifolds \cite{QBPASMF}.

A geometrical approach to study the evolution of multilevel quantum system
(qudit) was developed in \cite{QBPASMF,OCGQC, GAQCLB, QCAG, QGDM, GQCQ}. In \cite{OCGQC, GAQCLB, QCAG, QGDM} it was shown that
the problem of finding the quantum circuit of a unitary operations which provides time-optimal evolution on a system of qubits
is closely related to the problem of finding the minimal distance between two points on the Riemannian metric. A
similar problem was considered for the case of $n$ qutrits in
\cite{GQCQ}. The authors of that work showed that the
quantum gate complexity, which provides optimal evolution on a system of $n$ qutrits, is equivalent to the problem of finding
the shortest path between two points in a certain curved geometry of $SU\left(3^n\right)$.
The geometrical properties of some well known coherent state manifolds
were studied in \cite{RSMQS, CSRS}. More about geometry features of
multilevel quantum systems one can find in \cite{BSSQWF, BVQUDIT,GQM, BVNLS}.

In the previous paper \cite{prevpap} we proposed a two-step method for the preparation of an arbitrary quantum state on a two-spin system
represented by the isotropic Heisenberg Hamiltonian. In the present paper, we study the quantum evolution of a two-spin system with the isotropic Heisenberg
interaction in the external magnetic field (Section \ref{sec2}). In Section \ref{sec3} we show that this evolution happens on a two-parametric closed
manifold and calculate its Fubini-Study metric. It is shown that this manifold is a torus. The entanglement of the states which belong to this manifold is studied (Section \ref{sec4}). Conclusions are presented
in Section \ref{sec5}.

\section{The quantum evolution of a two-spin system \label{sec2}}

We consider a two-spin system represented by the isotropic Heisenberg Hamiltonian. The system is placed in the external magnetic field directed along the
$z$-axis. The Hamiltonian of the system is as follows
\begin{eqnarray}
H=H_{int}+H_{mf},
\label{form2}
\end{eqnarray}
with
\begin{eqnarray}
&&H_{int}=J\left(\mbox{\boldmath{$ \sigma$}}^1\mbox{\boldmath{$ \sigma$}}^2+1\right),
\label{form2_1}\\
&&H_{mf}=h_z\left(\sigma_z^{1}+\sigma_z^{2}\right),\label{form2_2}
\end{eqnarray}
where $\sigma_i^1=\sigma_i\otimes 1$,
$\sigma_i^2=1\otimes\sigma_i$ and $\sigma_i$ are the Pauli matrices,
$J$ is the interaction coupling, $h_z$ is proportional to the
strength of the magnetic field. Here $i=x,y,z$. It is worth noting that $H_{int}$ commutes with $H_{mf}$. This in turn means that Hamiltonian (\ref{form2}) and 
isotropic Heisenberg Hamiltonian (\ref{form2_1}) have a common set of eigenvectors
\begin{eqnarray}
&&\vert\uparrow\uparrow\rangle,\label{form3_1}\\
&&\vert\downarrow\downarrow\rangle,\label{form3_2}\\
&&\frac{1}{\sqrt{2}}\left(\vert\uparrow\downarrow\rangle + \vert\downarrow\uparrow\rangle\right),\label{form3_3}\\
&&\frac{1}{\sqrt{2}}\left(\vert\uparrow\downarrow\rangle - \vert\downarrow\uparrow\rangle\right).
\label{form3_4}
\end{eqnarray}
Hamiltonian $H_{int}$ has the three-fold degenerate eigenlevel $2J$ with eigenvectors (\ref{form3_1})-(\ref{form3_3}) (triplet state) and eigenlevel $-2J$ with
singlet state (\ref{form3_4}). Due to the external magnetic field, which splits energy levels, Hamiltonian (\ref{form2}) has four eigenvalues,
namely, $2\left(J+h_z\right)$, $2\left(J-h_z\right)$, $2J$, $-2J$ with the corresponding eigenvectors (\ref{form3_1})-(\ref{form3_4}).

Let us consider quantum evolution of a two-spin system with this Hamiltonian.
Taking into account that $H_{int}$ commutes with $H_{mf}$, we can represent the evolution operator in the following form
\begin{eqnarray}
U(t)=e^{-iH_{int}t}e^{-ih_z\sigma_z^1t}e^{-ih_z\sigma_z^2t},
\label{form4}
\end{eqnarray}
where
\begin{eqnarray}
e^{-iH_{int}t}=\cos{\left(2Jt\right)}-\frac{i}{2J}\sin{\left(2Jt\right)}H_{int}.\label{form4_1}
\end{eqnarray}
Here we use the fact that $H_{int}^{2}=(2J)^2$. We set $\hbar=1$, which means that the
energy is measured in frequency units. In the basis labelled by
$\vert\uparrow\uparrow\rangle$, $\vert\uparrow\downarrow\rangle$,
$\vert\downarrow\uparrow\rangle$ and
$\vert\downarrow\downarrow\rangle$, the evolution operator
$U(t)$ can be represented as
\begin{eqnarray}
U(t)=\left( \begin{array}{ccccc}
e^{-i2\left(h_z+J\right)t} & 0 & 0 & 0\\
0 & \cos\left(2J t\right) & -i\sin\left(2J t\right) & 0 \\
0 & -i\sin\left(2J t\right) & \cos\left(2J t\right) & 0 \\
0 & 0 & 0 & e^{i2\left(h_z-J\right)t}
\end{array}\right).
\label{form4_2}
\end{eqnarray}

Let us consider the evolution of a two-spin state having started from the initial state
\begin{eqnarray}
\vert\psi\rangle=a\vert\uparrow\uparrow\rangle+
b\vert\uparrow\downarrow\rangle+c\vert\downarrow\uparrow\rangle+d\vert\downarrow\downarrow\rangle,
\label{instate}
\end{eqnarray}
with parameters $a=a_i$, $b=b_i$, $c=c_i$ and $d=d_i$. The normalization condition is the following
$\vert a\vert^2+\vert b\vert^2+\vert c\vert^2+\vert d\vert^2=1$. The action of the evolution operator (\ref{form4})
on state (\ref{instate}) is as follows
\begin{eqnarray}
&&\vert\psi\left(\theta,\phi\right)\rangle=U(t)\vert\psi_i\rangle=a_ie^{-i\left(\phi+\theta\right)}\vert\uparrow\uparrow\rangle + \left(b_i\cos\theta-ic_i\sin\theta\right)\vert\uparrow\downarrow\rangle\nonumber\\
&& + \left(-ib_i\sin\theta+c_i\cos\theta\right)\vert\downarrow\uparrow\rangle + d_ie^{i\left(\phi-\theta\right)}\vert\downarrow\downarrow\rangle,
\label{form5}
\end{eqnarray}
where
\begin{eqnarray}
\theta=2Jt,\quad \phi=2h_zt.
\label{form5_1}
\end{eqnarray}
Note that this state is defined by two real independent parameters $\theta$ and $\phi$ which in turn are defined by the value of the magnetic field
$h_z$ and the period of evolution $t$. For any pre-defined set of values $\theta$ and $\phi$ there exists a set of values $h_z$ and $t$.
An arbitrary quantum state of two qubits contains six real parameters. Due to this fact, we cannot reach an arbitrary state of a
two-spin system, which is represented by Hamiltonian (\ref{form2}).

It is easy to see from (\ref{form5}) that the following equalities are satisfied
\begin{eqnarray}
&&\vert\psi(\theta+\pi,\phi)\rangle=-\vert\psi(\theta,\phi)\rangle,\label{form5_2}\\
&&\vert\psi(\theta,\phi+2\pi)\rangle=\vert\psi(\theta,\phi)\rangle.
\label{form5_3}
\end{eqnarray}
So, modulo a global phase, state $\vert\psi(\theta,\phi)\rangle$ is periodic with period $\pi$ for $\theta$ and with period $2\pi$ for $\phi$.
This means that parameters $\theta$ and $\phi$ belong to the intervals $\theta\in[0,\pi]$ and $\phi\in[0,2\pi]$, respectively.
In the next section in order to investigate the properties of the manifold, which contains all states achieved during the evolution of two spins, we consider the Fubini-Study metric.

\section{The Fubini-Study metric \label{sec3}}

The Fubini-Study metric is defined by the infinitesimal distance $ds$ between two neighbouring pure quantum states $\vert\psi(\xi^{\alpha})\rangle$ and
$\vert\psi(\xi^{\alpha}+d\xi^{\alpha})\rangle$ \cite{FSM, FSM2, FSM0, FSM1}
\begin{eqnarray}
ds^2=g_{\alpha\beta}d\xi^{\alpha} d\xi^{\beta},
\label{form10}
\end{eqnarray}
where $\xi^{\alpha}$ is a set of real parameters which define the state $\vert\psi(\xi^{\alpha})\rangle$. The components of the metric tensor
$g_{\alpha\beta}$ have the form
\begin{eqnarray}
g_{\alpha\beta}=\gamma^2\Re\left(\langle\psi_{\alpha}\vert\psi_{\beta}\rangle-\langle\psi_{\alpha}\vert\psi\rangle\langle\psi\vert\psi_{\beta}\rangle\right),
\label{form11}
\end{eqnarray}
where $\gamma$ is an arbitrary factor which is often chosen to have value $1$, $\sqrt{2}$ or $2$ and
\begin{eqnarray}
\vert\psi_{\alpha}\rangle=\frac{\partial}{\partial\xi^{\alpha}}\vert\psi\rangle.
\label{form12}
\end{eqnarray}
This metric can be obtained using expression for the Fubini-Study distance between two neighbouring pure states
$\vert\!\psi(\xi^{\alpha})\!\rangle$ and $\vert\!\psi(\xi^{\alpha}\!+\!d\xi^{\alpha})\!\rangle$ \cite{FSM,FSM3}
\begin{eqnarray}
ds^2=\gamma^2\left(1-\vert\langle\psi(\xi^{\alpha})\vert\psi(\xi^{\alpha}+d\xi^{\alpha})\rangle\vert^2\right),
\label{FSDNS}
\end{eqnarray}
when the state $\vert\psi(\xi^{\alpha}+d\xi^{\alpha})\rangle$ having expanded into a series up to the second term in
$d\xi^{\alpha}$.

Using expression (\ref{form11}) the metric of the manifold defined by a set of parameters $\xi^{\alpha}$ can be
obtained. For example, the metric tensor of the ground state of the quantum $XY$ chain in a transverse magnetic field
was calculated in \cite{FSM4}. In this case the authors find the metric tensor of the ground state manifold depending
on the exchange coupling and the magnetic field.

Let us calculate the metric of the manifold defined by state (\ref{form5}). This state is determined by two real parameters
$\theta$ and $\phi$. In order to find the components of the metric tensor (\ref{form11}) we calculate the following derivatives
\begin{eqnarray}
&&\vert \psi_{\theta}\rangle=-ia_ie^{-i\left(\phi+\theta\right)}\vert\uparrow\uparrow\rangle + \left(-b_i\sin\theta-ic_i\cos\theta\right)\vert\uparrow\downarrow\rangle\nonumber\\
&&+ \left(-ib_i\cos\theta-c_i\sin\theta\right)\vert\downarrow\uparrow\rangle -id_ie^{i\left(\phi-\theta\right)}\vert\downarrow\downarrow\rangle,\nonumber\\
&&\vert \psi_{\phi}\rangle=-i a_ie^{-i\left(\phi+\theta\right)}\vert\uparrow\uparrow\rangle +i d_ie^{i\left(\phi-\theta\right)}\vert\downarrow\downarrow\rangle,
\label{form16}
\end{eqnarray}
and then we obtain
\begin{eqnarray}
&&\langle\psi \vert \psi_{\theta}\rangle=-i\left[1-B\right],\quad \langle\psi \vert \psi_{\phi}\rangle=-iD,\nonumber\\
&&\langle\psi_{\theta} \vert \psi_{\theta}\rangle=1,\quad\langle\psi_{\phi} \vert \psi_{\phi}\rangle=A,\quad\langle\psi_{\phi} \vert \psi_{\theta}\rangle=D,
\label{form17}
\end{eqnarray}
where
\begin{eqnarray}
&&A=\vert a_i\vert^2+\vert d_i\vert^2,\quad B=\vert b_i-c_i\vert^2,\quad D=\vert a_i\vert^2-\vert d_i\vert^2.
\label{form171}
\end{eqnarray}
Substituting (\ref{form17}) into (\ref{form11}), we have
\begin{eqnarray}
ds^2=\gamma^2\left[B\left(2-B\right)(d\theta)^2+\left(A-D^2\right)(d\phi)^2+2BDd\theta d\phi\right].
\label{form18}
\end{eqnarray}
It is easy to show that the substitutions
\begin{eqnarray}
&&\phi' = \phi+k\theta,\\
&&\theta'=\theta
\label{subst}
\end{eqnarray}
transform this metric into diagonal form
\begin{eqnarray}
ds^2=\gamma^2\left[\frac{B\left(2A-2D^2-AB\right)}{A-D^2}(d\theta')^2+\left(A-D^2\right)(d\phi')^2\right],
\label{form19}
\end{eqnarray}
where $k=BD/\left(A-D^2\right)$.
As we can see from (\ref{form18}) or (\ref{form19}), the components of the metric tensor depend only on parameters which determine the initial states.
The state (\ref{form5}) with new parameters $\phi'$ and $\theta'$ takes the form
\begin{eqnarray}
&&\vert\psi\left(\theta',\phi'\right)\rangle=a_ie^{-i\left(\phi'+\left(1-k\right)\theta'\right)}\vert\uparrow\uparrow\rangle + \left(b_i\cos\theta'-ic_i\sin\theta'\right)\vert\uparrow\downarrow\rangle\nonumber\\
&& + \left(-ib_i\sin\theta'+c_i\cos\theta'\right)\vert\downarrow\uparrow\rangle + d_ie^{i\left(\phi'-\left(1+k\right)\theta'\right)}\vert\downarrow\downarrow\rangle.
\label{form20}
\end{eqnarray}
This state satisfies the following periodic conditions
\begin{eqnarray}
&&\vert\psi(\theta'+\pi,\phi'+k\pi)\rangle=-\vert\psi(\theta',\phi')\rangle,\label{form20_1}\\
&&\vert\psi(\theta',\phi'+2\pi)\rangle=\vert\psi(\theta',\phi')\rangle
\label{form20_2}
\end{eqnarray}
with respect to parameters $\theta'$ and $\phi'$.
Also, it is worth noting that the components of the metric tensor $g_{\theta'\theta'}$ and $g_{\phi'\phi'}$ have positive values.
Indeed, using notations (\ref{form171}) in (\ref{form19}), we obtain that
\begin{eqnarray}
&&B\left(2A-2D^2-AB\right)=(\vert a_i\vert^2+\vert d_i\vert^2)\vert b_i^2-c_i^2\vert^2+8\vert a_i\vert^2\vert d_i\vert^2\vert b_i-c_i\vert^2\ge 0,\nonumber\\
&&A-D^2=\vert a_i\vert^2(1-\vert a_i\vert^2)+\vert d_i\vert^2(1-\vert d_i\vert^2)+2\vert a_i\vert^2\vert d_i\vert^2\ge 0.
\label{form21}
\end{eqnarray}
So, we conclude that expression (\ref{form18}) or (\ref{form19}) defines the Euclidean metric.
The fact that components of the metric tensor do not depend on the parameters $\theta$ and $\phi$ means that expression (\ref{form18}) or (\ref{form19})
defines the metric of a flat manifold. Using this fact and the fact that parameters $\theta$ and $\phi$ are periodical, we conclude that this
manifold is a torus.

At the end of this section it is worth noting that the geometry of the manifold defined by expression
(\ref{form18}) or (\ref{form19}) depends on the parameters which determine the initial state. So, we obtain that the evolution of the
two-spin system happens on the one-dimensional manifold if the parameters of the initial state satisfy the condition $B=0$
or $A-D^2=0$. This condition is obtained from expression (\ref{form18}), when $g_{\theta\theta}=0$ or $g_{\phi\phi}=0$.
Let us study entanglement of the states which belong to this manifold.

\section{Entanglement on the torus\label{sec4}}

The implementation of different algorithms in quantum computation demands preparation of maximally entangled states.
For example, the realization of the simplest scheme of the quantum teleportation
of one qubit state requires preparation of EPR channel \cite{teleport}.

In \cite{ent0} it was considered the entanglement of multipartite system using geometry property of the Hilbert space.
Also the authors considered the case of the bipartite entanglement. They gave the definition of the concurrence for an arbitrary mixed state
using geometry language. In this section we investigate the degree of concurrence of the states which belong to the manifold
defined by metric (\ref{form18}).

The degree of entanglement of two spin-system can be determined by the concurrence \cite{ent1}, \cite{ent2}
\begin{eqnarray}
C=2\vert ad-bc\vert,
\label{form32}
\end{eqnarray}
where parameters $a$, $b$, $c$ and $d$ are defined by expression (\ref{instate}).
Using this definition, we obtain that the concurrence in state (\ref{form5}) is as follows
\begin{eqnarray}
C=2\left\vert a_id_i e^{-i2\theta}-\left(b_ic_i\cos 2\theta-\frac{i}{2}\left({b_i}^2+{c_i}^2\right)\sin 2\theta \right)\right\vert.
\label{form33}
\end{eqnarray}
It is easy to see that the entanglement of state (\ref{form5}) depends only on parameter $\theta$ which contains the interaction between spins
and is independent of parameter $\phi$ which contains the value of the magnetic field. This is because the action of the magnetic field is
given by unitary operators that define the evolution of each spin separately and do not change the entanglement of the spin system.
While the interaction between spins is defined by unitary operator which describes the evolution of the two spins together. This, in turn,
leads to a change in the entanglement of two spins. From equation (\ref{form33}) it follows that for a particular value of
$\theta$ we can select the curves on our manifold with a constant entanglement. From expression (\ref{form18}) or (\ref{form19})
we obtain that these curves are circles with radii depending on parameters of initial states as follows
\begin{eqnarray}
R=\gamma\sqrt{A-D^2},
\label{form331}
\end{eqnarray}
where $A$ and $D$ are defined by (\ref{form171}).
Also, it is worth noting if the initial state is disentangled ($C=0$) than expression (\ref{form33}) takes the form
\begin{eqnarray}
C=\vert b_i-c_i\vert^2\vert \sin2\theta\vert.
\label{form332}
\end{eqnarray}
This is due to the fact that $a_id_i=b_ic_i$. As we can see in the case of $\theta=\pi/4$ we obtain the value of the maximal entanglement.
This value depends on the parameters $b_i$ and $c_i$ that determine the initial state.
For instance, in the case of $b_i=1$ and $c_i=0$, which corresponds to the initial state $\vert\uparrow\downarrow\rangle$,
we obtain the maximally entangled state with $C=1$. Let us consider
another example when parameters $b_i$ and $c_i$ are the same, $b_i=c_i$. In this case we never achieve an entangled state. This is because the following initial state
$\vert\psi_i\rangle=a_i\vert\uparrow\uparrow\rangle+b_i\left(\vert\uparrow\downarrow\rangle+\vert\downarrow\uparrow\rangle\right)+d_i\vert\downarrow\downarrow\rangle$
is an eigenstate of the two-spin system (\ref{form2_1}).

Let us apply the above results to the disentangled initial state which has the following form
\begin{eqnarray}
\vert\psi_i\rangle=\vert+-\rangle,
\label{form34}
\end{eqnarray}
where $\vert+\rangle=\cos\frac{\chi}{2}\vert\uparrow\rangle+\sin\frac{\chi}{2}e^{i\gamma}\vert\downarrow\rangle$,
$\vert-\rangle=-\sin\frac{\chi}{2}\vert\uparrow\rangle+\cos\frac{\chi}{2}e^{i\gamma}\vert\downarrow\rangle$ are the
eigenstates of the operator of projection of spin-$\frac{1}{2}$ on the direction defined by the unit vector ${\bf n}$. The vector ${\bf n}$ is represented by the spherical coordinates as follows ${\bf n}=\left(\sin\chi\cos\gamma,\sin\chi\sin\gamma,\cos\chi\right)$,
where $\chi\in[0,\pi]$ and $\gamma\in[0,2\pi]$ are the polar and azimuthal angles, respectively. It is also important to note that state (\ref{form34}) is the eigenstate of the system of
two spins in the magnetic field directed along the unit vector ${\bf n}$. Substituting parameters of states (\ref{form34})
in (\ref{form18}) we obtain that in this case the evolution happens on a torus with the metric defined by
\begin{eqnarray}
ds^2=\gamma^2\left((d\theta)^2+\frac{1}{2}\sin^2\chi(d\phi)^2\right).
\label{form35}
\end{eqnarray}
From equation (\ref{form332}) we obtain that the concurrence of the states on the manifold defined by (\ref{form35})
depends on the parameter $\theta$ as follows
\begin{eqnarray}
C=\vert\sin2\theta\vert.
\label{form36}
\end{eqnarray}
It is easy to see that the condition $\theta=\frac{\pi}{4}$ corresponds to the maximally entangled state
\begin{eqnarray}
&&\vert\psi\left(\theta,\phi\right)\rangle=-\cos\frac{\chi}{2}\sin\frac{\chi}{2}e^{-i\left(\phi+\frac{\pi}{4}\right)}\vert\uparrow\uparrow\rangle + \frac{1}{\sqrt{2}}\left(\cos^2\frac{\chi}{2}+i\sin^2\frac{\chi}{2}\right)e^{i\gamma}\vert\uparrow\downarrow\rangle\nonumber\\
&& + \frac{1}{\sqrt{2}}\left(i\cos^2\frac{\chi}{2}+\sin^2\frac{\chi}{2}\right)e^{i(\gamma+\pi)}\vert\downarrow\uparrow\rangle + \cos\frac{\chi}{2}\sin\frac{\chi}{2}e^{i2\gamma}e^{i\left(\phi-\frac{\pi}{4}\right)}\vert\downarrow\downarrow\rangle.
\label{form37}
\end{eqnarray}
Taking into account (\ref{form5_1}), we conclude that the system having started from the initial state (\ref{form34}) can achieve state (\ref{form37}) during the period of time
\begin{eqnarray}
t=\frac{\pi}{8J}.
\label{time}
\end{eqnarray}
If the parameter $\chi=0$, which modulo a global phase corresponds
to the case of initial state $\vert\uparrow\downarrow\rangle$, then we obtain the following final state
\begin{eqnarray}
\vert\psi\left(\theta,\phi\right)\rangle=\frac{1}{\sqrt{2}}\vert\uparrow\downarrow\rangle - \frac{i}{\sqrt{2}}\vert\downarrow\uparrow\rangle.
\label{form38}
\end{eqnarray}
Evolution between the initial state $\vert\uparrow\downarrow\rangle$ and final one (\ref{form38}) happens along the curve which is a circle with the radius $\gamma$.
It is easy to see if we put $\chi=0$ in expression (\ref{form35}).

Finally, it should be noted that in the case of the initial state
$\vert++\rangle$ or $\vert--\rangle$ the evolution happens on the manifold which is the circle with the radius
$\gamma\sin\chi/\sqrt{2}$. Also, it is easy to see that all the states
belonging to this circle are disentangled because the initial state is an eigenstate of Hamiltonian (\ref{form2_1}).

\section{Conclusion \label{sec5}}

We considered the quantum system of two spins represented by the isotropic Heisenberg Hamiltonian.
The quantum evolution of such a system placed in an external magnetic field which is directed along the $z$-axis was studied.
This evolution is defined by two real parameters, namely, the period of time of evolution and the value of the magnetic field.
The evolution of the two-spin system is periodic with respect to these parameters. Therefore, we concluded that this evolution happens
on a two-parametric closed manifold.
We calculated the Fubini-Study metric of this manifold and showed that it describes a flat manifold. Using this fact
and the fact that parameters which define this manifold are periodic we concluded that it is a torus. Finally, the entanglement of the states belonging to this manifold
was investigated. We found that the curves of constant entanglement on the manifold are circles.
Also we showed that the evolution between the disentangled and maximally entangled states happens on a torus.

\section{Acknowledgment}

First of all, the authors would like to thank Yuri Krynytskyi for useful
comments. Also the authors would like to thank Dr. Andrij Rovenchak and Khrystyna Gnatenko. The paper is based on the research provided
by the grant support of the State Fund for Fundamental Research of Ukraine, Project F-64/41-2015 (No. 0115U004838).


\begin{thebibliography}{99}
\bibitem{GFofQM} Dariusz Chru\'{s}ci\'{n}ski, J. Phys: Conference Series \textbf{30}, 9 (2006).
\bibitem{GAQMQE} Abhay Ashtekar and Troy A. Schilling, gr-qc/9706069.
\bibitem{AGAtoQM} J. Anandan, Foundation of Physics \textbf{21}, 1265 (1991).
\bibitem{EMCCR} E. Ercolessi, G. Marmo and G. Morandi, Riv.Nuovo Cim. {\bf 033}, 401 (2010).
\bibitem{FSM} V. M. Tkachuk, {\it Fundamental problems of quantum mechanic} (Lviv: Ivan Franko National University of Lviv) (2011). [in Ukrainian]
\bibitem{FSM2} I. Bengtsson and K. \.Zyczkowski, {\it Geometry of quantum states}, (New York: Cambridge University press) (2006).
\bibitem{ref1} Carl M. Bender, Dorje C. Brody, Hugh F. Jones and Bernhard K. Meister, Phys. Rev. Lett. \textbf{98}, 040403 (2007).
\bibitem{ref2} Carl M. Bender, and Dorje C. Brody, Time in Quantum Mechanics --
\textbf{2}, Lecture Notes in Physics, \textbf{789}, 341 (Berlin: Springer) (2010).
\bibitem{ref3} Carl M. Bender, SIGMA \textbf{3}, 126 (2007).
\bibitem{OHfST} Dorje C. Brody, and Daniel W. Hook, J. Phys. A \textbf{39}, L167 (2006).
\bibitem{ZNP1} Benjamin Russell, Susan Stepney, Phys. Rev. A {\bf 90}, 012303 (2014).
\bibitem{ZNP2} Benjamin Russell, Susan Stepney, J. Phys. A, {\bf 48}, 115303 (2015).
\bibitem{ZNP3} Dorje C. Brody, Gary W. Gibbons, David M. Meier, New J. Phys. {\bf 17}, 033048 (2015).
\bibitem{ZNP4} Dorje C. Brody, David M. Meier, Phys. Rev. Lett. {\bf 114}, 100502 (2015).
\bibitem{ZNP5} Dorje C. Brody, David M. Meier, J. Phys. A {\bf 48}, 055302 (2015).
\bibitem{QBPASMF} A. R. Kuzmak, V. M. Tkachuk, Phys. Lett. A \textbf{379}, 1233 (2015).
\bibitem{OCGQC} M. A. Nielsen, M. R. Dowling, M. Gu and A. C. Doherty, Phys. Rev. A {\bf 73}, 062323 (2006).
\bibitem{GAQCLB} M. A. Nielsen, Quant. Inform. Comput. {\bf 6}, 213 (2006).
\bibitem{QCAG} M. A. Nielsen, M. R. Dowling, M. Gu and A. C. Doherty, Science {\bf 311}, 1133 (2006).
\bibitem{QGDM} Navin Khaneja, Bj\"orn Heitmann, Andreas Sp\"orl, Haidong Yuan, Thomas Schulte-Herbr\"uggen and Steffen J. Glaser, arXiv:quant-ph/0605071 (2006).
\bibitem{GQCQ} Bin Li, Zu-Huan Yu and Shao-Ming Fei, Nature Scientific Reports {\bf 3}, 2594 (2013).
\bibitem{RSMQS} J. P. Provost and G. Vallee, Commun. Math. Phys. {\bf 76}, 289 (1980).
\bibitem{CSRS} Dorje C. Brody and Eva-Maria Graefe, J. Phys. A {\bf 43} 255205 (2010).
\bibitem{BSSQWF} Surajit Sen, Mihir Ranjan Nath, Tushar Kanti Dey and Gautam Gangopadhyay, Annals of Physics {\bf 327}, 224 (2012).
\bibitem{BVQUDIT} Reinhold A. Bertlmann and Philipp Krammer, J. Phys. A {\bf 41} 235303 (2008).
\bibitem{GQM} Dorje C. Brody and Lane P. Hughston, Journal of Geometry and Physics {\bf 38}, 19 (2001).
\bibitem{BVNLS} Gen Kimura, Phys. Lett. A {\bf 314}, 339 (2003).
\bibitem{prevpap} A. R. Kuzmak, V. M. Tkachuk, Phys. Lett. A \textbf{378}, 1469 (2014).
\bibitem{FSM0} D. N. Page, Phys. Rev. A {\bf 36}, 3479 (1987).
\bibitem{FSM1} S. Kobayashi, K. Nomizu, {\it Fundations of Differential Geometry}, Vol. 2, Wiley, New York, (1969).
\bibitem{FSM3} Sumiyoshi Abe, Phys. Rev. A {\bf 46}, 1667 (1992).
\bibitem{FSM4} M. Kolodrubetz, V. Gritsev and A. Polkovnikov, Phys. Rev. B {\bf 88}, 064304 (2013).
\bibitem{teleport} Charles H. Bennett, Gilles Brassard, Claude Cr\'epeau, Richard Jozsa, Asher Peres, William K. Wootters,
Phys. Rev. Lett. {\bf 70}, 1895 (1993).
\bibitem{ent0} J. Grabowski, M. Kus, G. Marmo, Open Sys. and Information Dyn. {\bf 13}, 343 (2006).
\bibitem{ent1} W. K. Wootters, Phys. Rev. Lett. {\bf 80}, 2245 (1998).
\bibitem{ent2} S. Hill and W. K. Wootters, Phys. Rev. Lett. {\bf 78}, 5022 (1997).
\end{thebibliography}
\end{document}